\documentclass[aps,showpacs,amssymb,prl,amsmath,prl,twocolumn,reprint]{revtex4}
\usepackage{amsmath,latexsym,amssymb}
\usepackage{graphicx}
\usepackage{amsmath}
\usepackage{latexsym}
\usepackage{amssymb}
\usepackage{bm}

\begin{document}

\title{Cancellation of Collisional Frequency Shifts in Optical Lattice Clocks with Rabi Spectroscopy}
\author{Sangkyung Lee, Chang Yong Park, Won-Kyu Lee, and Dai-Hyuk Yu}

\affiliation{Korea Research Institute of Standards and Science, Daejeon 305-340, Republic of Korea}

\pacs{42.62.Eh, 42.62.Fi, 32.70.Jz, 34.50.-s}

\date{\today}

\begin{abstract}
We analyze both the s- and p-wave collision induced frequency shifts and propose a over-$\pi$ pulse scheme to cancel the shifts in optical lattice clocks interrogated by a Rabi pulse. The collisional frequency shifts are analytically solved as a function of the pulse area and the inhomogeneity of the Rabi frequencies. Experimentally measured collisional frequency shifts in an Yb optical lattice clock are in good agreement with the analytical calculations. Based on our analysis, the over-$\pi$ pulse combined with a small inhomogeneity below 0.1 allows a fractional uncertainty on a level of $10^{-18}$ in both Sr and Yb optical lattice clocks by canceling the collisional frequency shift. 
\end{abstract}

\maketitle

Ultracold atoms trapped in an optical lattice have served as ample frameworks for high-precision atomic clocks~\cite{JYe,NHinkley,BJBloom,CWChou}, quantum information processing~\cite{DHayes, POSchmidt}, and for quantum simulators with Bose-Einstein condensates~\cite{JStruck, IBloch}. Many atoms in the same optical lattice site inevitably lead to collisions among atoms; therefore, efforts to understand  and control the collision dynamics are crucial to improve performance levels. For optical lattice clocks, the simultaneous interrogation of many atoms is a key feature in contrast to ion clocks, and a reduction of the collision shift is one of the most important factors to reach relative frequency uncertainty level of $10^{-18}$. To overcome collisional shifts, the suppression of these collisions by strong atom-atom interactions in a Sr 2D optical lattice~\cite{MDSwallows1} and precise measurements of the collision shift per atom number in a Sr 1D optical lattice with Rabi spectroscopy~\cite{TLNicholson1} were reported. Collision-shift cancellation has also been demonstrated with Ramsey spectroscopy~\cite{NDLemke1, ADLudlow1} and with Rabi spectroscopy~\cite{GKCampbell, AMRey1, KGibble1}.  Recently it has been shown that the inhomogeneity of Rabi frequencies allows s-wave collisions~\cite{GKCampbell,ZYu, KGibble1, MDSwallows1} and that p-wave collisions are even dominant in Yb and Sr lattice clocks~\cite{NDLemke1, MJMartin}. However, in Rabi spectroscopy, the cancellation of the shift were investigated by only terms of s-wave collisions in the case that the initial state is the excited state. The p-wave collisions gives the dramatic difference of the shift with respect to the s-wave collision case, especially when the initial state is the ground state. Nowadays the Rabi line shape distortion in the p-wave dominant case has been studied in aspect of observing many-body interactions~\cite{MJMartin}. In contrast to Ramsey spectroscopy, the cancellation of the p-wave collision induced shift, which is important in optical lattice clocks, has not been explored and analytically analyzed in Rabi spectroscopy.    

In this Letter, we analyze both s- and p-wave cold collisions and propose a over-$\pi$ pulse interrogation scheme for canceling the collision shift in optical lattice clocks using Rabi spectroscopy. We apply our analysis to the experimentally measured collisional frequency shift in an Yb optical lattice clock. Although shift and uncertainty on the $10^{-17}$ level are reported due to operation at slightly off of collision-shift-free conditions, it is shown that shift and uncertainty on the $10^{-18}$ level can be reached under experimentally achievable conditions, such as a small inhomogeneity below 0.1 in the Rabi frequencies of trapped atoms and a small degree of lock servo-fluctuation ($\pm 1~\%$), for both Yb and Sr lattice clocks when using an optimum over-$\pi$ pulse.

\begin{figure}
\includegraphics[width=0.40\textwidth]{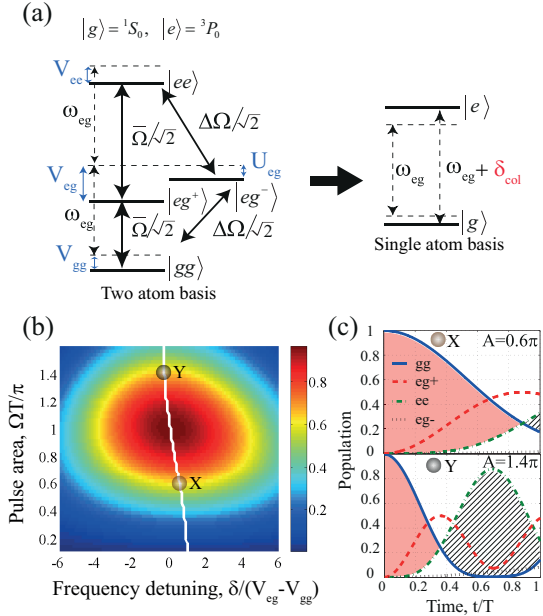}
\caption{(Color online) (a) (Left) Energy level diagram of ultracold two-atom collisions on triplet and singlet bases.  (Right) Energy level diagram on a single-atom basis. 
(b) Excited population as a function of the laser frequency detuning and pulse area, obtained by projecting the singlet and triplet populations onto the single-atom basis. The white solid line depicts the peak of the transition spectrum. (c) Time evolution of the probabilities of singlet and triplet states. The red shaded (or hatched) region represents the positive (or negative) time-averaged population difference, $\int_0^T{N_g(t) - N_e(t) dt/T}>0$ (or $\int_0^T{N_g(t) - N_e(t) dt/T}<0$). }
\label{fig1}
\end{figure}

In actual 1D optical lattice cases, more than two atoms exist in a disk. Although a many-body Hamiltonian is required to describe the 1D case~\cite{AMRey1,MJMartin,AMRey2}, the simple two-atom four-level Hamiltonian shown in Eq.~\eqref{eq1} with effective collision-interaction frequencies can predict many-atom cases in the weak atom-atom interaction regime by adopting a mean field picture~\cite{NDLemke1}. In order to calculate the collisional frequency shifts in Rabi spectroscopy, we solve the Schr\"{o}dinger equation with the four-level Hamiltonian on a two-atom basis, in this case singlet($\left|eg^-\right>=\frac{1}{\sqrt{2}}\left[\left|eg\right>-\left|ge\right>\right]$) and triplet states($\left|gg\right>$, $\left|ee\right>$, and  $\left|eg^+\right>=\frac{1}{\sqrt{2}}\left[\left|eg\right>+\left|ge\right>\right]$), 
\begin{equation}
\frac{H}{\hbar} = 
\begin{pmatrix}
 \delta+V_{gg} & 0 &\bar{\Omega}/ \sqrt{2} &\Delta \Omega/ \sqrt{2} \\
0 & -\delta+V_{ee}&\bar{\Omega}/\sqrt{2} &-\Delta\Omega/\sqrt{2} \\ 
\bar{\Omega}/\sqrt{2} & \bar{\Omega}/\sqrt{2} & V_{eg} & 0 \\
\Delta\Omega/\sqrt{2} & -\Delta\Omega/\sqrt{2} & 0 &U_{eg} 
\end{pmatrix},
\label{eq1}
\end{equation}
where $V_{gg}$ and $V_{ee}$, and $V_{eg}$ and $U_{eg}$ correspond to the frequencies characterizing the p-wave collision shifts of the triplet states and the s-wave collision shifts of the singlet state, respectively. These frequency shifts are determined by the scattering lengths and the atomic density. $\delta$ denotes the laser detuning with respect to the bare atom resonance frequency, $\omega_{eg}$. $\dot{\phi}$ is the frequency error of the clock laser. $\bar{\Omega}$ is the thermally averaged Rabi frequency of atoms in vibrational states in an optical lattice and $\Delta \Omega$ is the RMS spread in the Rabi frequency, $\Delta\Omega = \sqrt{\left<\Omega^2_{\vec{n}}\right>_T-\bar{\Omega}^2}$, where $\left<\Omega^2_{\vec{n}}\right>_T$ is the thermal average of the square of the Rabi frequencies. We define the inhomogeneity of the Rabi frequency as $\gamma = \Delta\Omega/\bar{\Omega}$. 

The populations of the triplet and singlet states are projected to a single-atom basis, $\left|e\right>$ and $\left|g\right>$, as shown in Fig.~\ref{fig1}-(a), to obtain the e-g transition spectrum which is the expectation value of the projection operator, $\hat{P}_{e} = \left|ee\right>\left<ee\right|+\frac{1}{2}\left|eg^{+}\right>\left<eg^{+}\right|$. The collisional frequency shift of the clock transition is determined by measuring the center of the e-g transition spectrum, as shown in Fig.~\ref{fig1}-(b). As intuitively expected and shown in more detail below, the collisional frequency shift depends mainly on the time-averaged population difference ({\it i.e.}, $\int_0^{T}{N_g(t) - N_e(t)}dt/T$, where $T$ is the pulse duration of a Rabi pulse). This is shown in Figs.~\ref{fig1}-(b) and (c), where the X point (or Y point) with a positive (or negative) shift shows a positive (or negative) time-averaged population difference. 

With the help of the second-order perturbation theory under the condition of $\bar{\Omega} \gg V_{ij=e,g}$, $U_{eg}$, $\delta$, we can obtain the analytical formula of the collisional frequency shift as a function of the pulse area $A=\bar{\Omega} T$. The collisional frequency shift is represented by the sum of a homogeneous p-wave, an inhomogeneous p-wave, and the s-wave collisional frequency shifts, $\delta_{col} = \delta^{(p)}_{hcol}+\delta^{(p)}_{ihcol}+\delta^{(s)}_{ihcol}$. 

\begin{figure}
\includegraphics[width=0.5\textwidth]{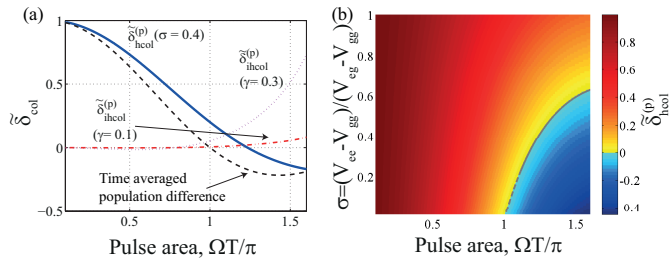}
\centering
\caption{(Color online) (a) Normalized {\it HPCFS} and {\it {\it IPCFS}} as a function of the pulse area. The blue solid line and the black dashed line depict the {\it HPCFS}, $\tilde{\delta}_{hcol}^{(p)}$, when $\sigma = 0.4$ and the time-averaged population difference, respectivley. The red dashed-dotted line and the purple dotted line represent the {\it {\it IPCFS}}, $\tilde{\delta}_{ihcol}^{(p)}$, when $\gamma = 0.1$ and $\gamma =0.3$, respectively. 
 (b) Normalized collisional frequency shift as a function of $\sigma$ and the pulse area. 
 }
\label{fig_cc}
\end{figure}

The homogeneous p-wave collisional frequency shift({\it HPCFS}) is the dominant term among the three types of collisional frequency shift. It comes from the g-e population difference due to the time-varying probabilities of the triplet states assisted by $\bar{\Omega}$. It can be written as 
\begin{equation}
\delta^{(p)}_{hcol} \approx -\frac{\sin A}{A}(V_{eg}-V_{gg})\alpha(A)+(V_{ee}-V_{gg})\beta(A),
\label{eqhcol}
\end{equation}
where $\alpha(A) = \left(-1+\frac{\sin A}{A}\right)/(2\xi(A))$, $\beta(A) = [\frac{2\sin^2{(A/2)}}{A^2}+\frac{\sin A}{4A}(\frac{\sin A}{A}-3)]/\xi(A)$, and $\xi(A) = 4\frac{\sin^2(A/2)}{A^2}-\frac{\sin A }{A}$. Because the time-averaged population difference is given by $\int_0^{T} {N_g(t) - N_e(t)}dt/T \simeq {\sin A}/{A }$ in the perturbative regime, it is revealed that the homogeneous p-wave collisional frequency shift depends on the time averaged population difference, $\delta^{(p)}_{hcol} \approx C_1 + C_2\int_{0}^{T}{N_g(t)-N_e(t)}dt/T$, as shown in Fig.~\ref{fig_cc}-(a).

The inhomogeneous p-wave collisional frequency shift({\it IPCFS}) is determined by the g-e population difference resulting from the leakage of the triplet-state population to the singlet state due to $\Delta\Omega$. It can be written as
\begin{align}
\delta^{(p)}_{ihcol} &\approx \gamma^2[\theta(A)(V_{eg}-V_{gg})+ \eta(A)(V_{ee}-V_{gg}) \\ \nonumber &-\chi(A)V_{gg}],
\label{eqihcol}
\end{align}
where $\theta(A) = [-3+4A^2+2A^2\cos A + 3\cos 2A - A^3 \sin A]/(4A^2\xi(A))$, $\eta(A) = [75-6A^2+24(-3+A^2)\cos A - 3\cos(2A) - 60A\sin A +7A^3\sin A]/(24A^2\xi(A))$, and $\chi(A) = [\cos^2{\frac{A}{2}}-(\frac{\sin A}{A} + \frac{A^2}{6})\frac{\sin A}{A}]/\xi(A)$. The {\it {\it IPCFS}} increases with the pulse area, dominated by $\theta(A)$. 

The singlet-state $\left|eg^-\right>$ populated according to the nonzero $\Delta\Omega$ undergoes an s-wave collision. The s-wave collisional frequency shift({\it SCFS}) is written as
\begin{equation}
\delta^{(s)}_{ihcol} \approx \gamma^2\chi(A)U_{eg}.
\label{eqsihcol}
\end{equation}
The {\it SCFS} is 0 at $A=\pi$, which indicates that the time-averaged population difference is 0~\cite{AMRey1,KGibble1}. This is the smallest term among the three types of collision shifts. 

We normalize the collisional frequency shift to $\tilde{\delta}_{col} = \frac{\delta_{col}}{(V_{eg}-V_{gg})}$ to verify the zero-crossing behavior of the collisional frequency shift independent of $V_{eg}-V_{gg}$. Each normalized collision shift is plotted in Fig.~\ref{fig_cc}-(a). The negative {\it HPCFS} in the over-$\pi$ pulse area regime is the heart of the cancellation of the collisional frequency shift because the {\it {\it IPCFS}} is always positive (see Fig.~\ref{fig_cc}-(a)). To realize the cancellation of the negative {\it HPCFS} and the positive {\it {\it IPCFS}}, a sufficiently small inhomogeneity is required.

The {\it HPCFS} can be positive in the entire range of the pulse area in the case of a large $V_{ee}-V_{gg}$ values; hence the cancellation of the collisional frequency shifts is not possible in such a case. To check the necessary condition of $V_{ee}-V_{gg}$ for cancellation, {\it i.e.}, $\delta_{hcol}^{(p)}\leq 0$, the {\it HPCFS} is calculated as a function of the pulse area and the ratio between the collision interaction frequencies $\sigma = \frac{V_{ee}-V_{gg}}{V_{eg}-V_{gg}}$, as shown in Fig.~\ref{fig_cc}-(b). 
The small value of $\sigma<0.63$ is a necessary condition for the cancellation of the collisional frequency shift. Fortunately, the cancellation of the collisional frequency shift is possible in both the Sr $(\sigma \simeq 0.4)$ and Yb $(\sigma \simeq 0.1)$ optical lattice clocks~\cite{NDLemke1, AMRey2}.  

In the beyond-perturbation regime, {\it i.e.,} $\Omega < V_{ij=e,g}$, the Rabi spectrum is distorted and the shift starts to stray from the mean-field approach~\cite{AMRey2}. The shift is not clearly defined in the distorted Rabi line shape. In an atomic clock based on the Rabi spectroscopy, its operation in the perturbative regime is appropriate.

\begin{figure}
\includegraphics[width=0.48\textwidth]{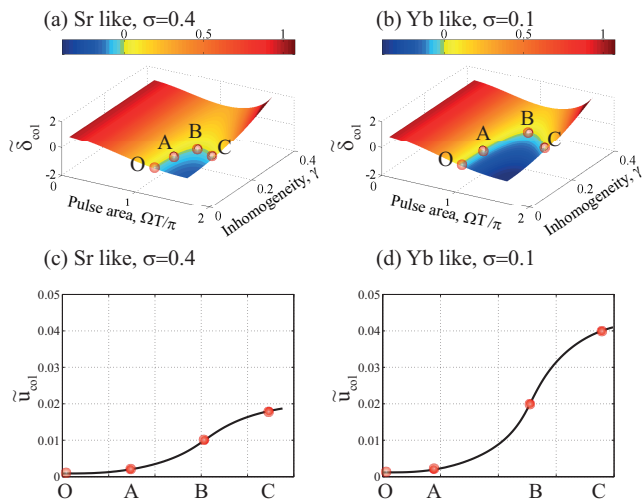}
\centering
\caption{(Color online) (a) and (b) Normalized collisional frequency shift $\tilde{\delta}_{col}$ as a function of the pulse area and the inhomogeneity in Sr and Yb. (c) and (d) Normalized uncertainty $\tilde{u}_{col}$ along with the zero shift curve in Sr and Yb. It is assumed that $\delta A = 0.1~\%$ and $\delta\gamma = 5~\%$. }
\label{fig_sryb}
\end{figure}    
 
 We calculate the total collisional frequency shift as a function of the pulse area and the inhomogeneity in both Yb and Sr atoms using Eqs.~\eqref{eqhcol}-\eqref{eqsihcol}. Figures~\ref{fig_sryb}-(a) and (b) show the existence of the zero-collisional frequency shift in the over-$\pi$ pulse regime.
 
Figures~\ref{fig_sryb}-(c) and (d) show the normalized uncertainties in the $\tilde{\delta}_{col}(A,\gamma) = 0$ curves of Sr and Yb. Fluctuations in the atomic density, pulse area, and inhomogeneity are the main origins of the residual collisional frequency shift change. The normalized deviations of the residual collisional frequency shift changes({\it i.e.} normalized uncertainties) are calculated by $\tilde{u}_{col} = \sqrt{(\tilde{\delta}_{col}\frac{\delta\rho}{\rho})^2+(\frac{\partial \tilde{\delta}_{col}}{\partial A}\delta A)^2+(\frac{\partial \tilde{\delta}_{col}}{\partial \gamma}\delta\gamma)^2}$~\cite{A0}. In the collision shift-free $\tilde{\delta}_{col}(A,\gamma)=0$ curve, the uncertainty stemming from the fluctuation of the atomic density, $\tilde{\delta}_{col}\frac{\delta\rho}{\rho}$, disappears. Table~\ref{table1} shows the conditions for $\tilde{u}_{col} \leq 2\times 10^{-3}$, as depicted by the O-A line in Fig.~\ref{fig_sryb}. $V_{eg}-V_{gg}$ under the Hz level can provide $5\times 10^{-18}$ uncertainty in this condition. The atomic density should be on the order of $10^{10}~$/cm$^3$; this can be obtained by the large volume optical cavity of an optical lattice without sacrificing the S/N ratio. 

  \begin{table}[ht]
 \caption{Collision-shift cancellation condition for $\tilde{u}_{col} \leq 2\times 10^{-3}$.  }
 \centering
 \begin{tabular}{c  c c  c c }
 \hline
 \hline
 Atom & $\sigma$ & $A$ & $\gamma$    &  $\left|V_{eg}-V_{gg}\right|$ for a $5\times 10^{-18}$  \\
 & & &    & (Hz)\\
 \hline
 
 Sr&$0.4$ & $A\leq1.25\pi$ & $\gamma\leq0.09$ & $\leq1.07$\\

 Yb& $0.1$ &$A\leq1.05\pi$& $\gamma\leq0.11$  & $\leq1.30$  \\

 \hline
 \hline
 \end{tabular}
 \label{table1}
 \end{table}

To apply our analysis to experiments, we measure the collisional frequency shifts of an $^{171}$Yb optical lattice clock as a function of the pulse area in two inhomogeneity cases, $\gamma = 0.23$ and $\gamma = 0.15$, the latter being the lowest in our current experimental setup. The details of our Yb optical lattice clock are described in the literature~\cite{CYPark1, DHYu1}. The potential depth of the vertically oriented 1D optical lattice is $U_0 = 185E_r$ and the trap frequencies are $\nu_z \simeq 55$ kHz and $\nu_r \simeq 202$ Hz, as measured by the sideband spectra~\cite{SBlatt1}. 
Nearly $50000$ atoms are trapped in the optical lattice. 
The atom density here is roughly $\rho_0 = 4.3\times10^{10}$ atoms/cm$^3$ where $V_{ij=e,g}$ and $U_{eg}$ are sub-Hz levels. To measure the collisional frequency shift at the given atom density, the interlacing between the high ($N_0$) and low atom numbers($0.57N_0$) is repeated. The final excited fraction is fixed at 90\%($\pm 6\%$) of the maximum excited fraction by feedback to the computer-controlled AOM for the clock laser (a clock laser lock servo)~\cite{A-1}. The atomic sample is spin-polarized ($m_F = 1/2$ or $-1/2$) by properly detuned 556 nm optical pumping in the presence of an 8.8 Gauss bias magnetic field. The inhomogeneity of our system is 0.23 at a temperature of $4.9(0.4)~\mu$K. To decrease the inhomogeneity, atoms in higher vibrational states of the optical lattice have to be blown out~\cite{SFalke}. We drop the intensity of the lattice laser to 53\% level, at which the potential depth reaches $98E_r$. After 20 ms, the intensity of the lattice laser is increased to the original level(the intensity down-up method). The inhomogeneity is lowered to 0.15, where the temperature is $3.9(0.3)~\mu$K. The pulse area is controlled by the power of the clock laser at a fixed short pulse duration of $T=10$ ms.

\begin{figure}
\includegraphics[width=0.35\textwidth]{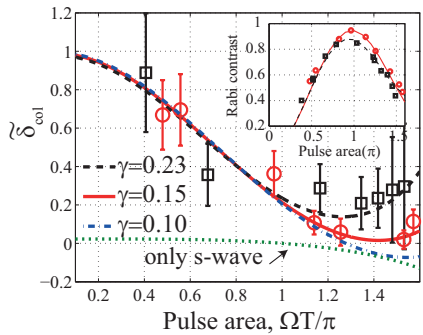}
\centering
\caption{(Color online) Normalized collisional frequency shifts $\tilde{\delta}_{col}$ as a function of the pulse area in Yb. The black dashed line ($\gamma = 0.23$), the red solid line ($\gamma = 0.15$), and the blue dashed-dotted line ($\gamma = 0.10$) are obtained from the analytical formula. The black squares and the red circles depict the experimental data at $\gamma = 0.23$ and $\gamma = 0.15$, respectively. The only s-wave collision case, $\delta^{(s)}_{ihcol}/U_{eg}$ is plotted as the green dotted line. The inset shows the Rabi contrasts at $\gamma = 0.23$ (black dashed line) and $\gamma = 0.15$ (red solid line). }
\label{fig2}
\end{figure}

Figure \ref{fig2} shows the experimental and calculated normalized collisional frequency shifts as a function of the pulse area. The analytical solutions show good agreement with the experimental data. The only s-wave collision shift (the green dotted line) does not explain the measured shift and therefore we confirm that the p-wave collisions are dominant, which is consistent with the result of Ramsey spectroscopy~\cite{NDLemke1}. Although the measured shift does not cross the zero shift mark due to the inhomogeneity limit of 0.15 in our experimental setup, the calculation proves that over-$\pi$ pulse cancellation is possible at an inhomogeneity level of less than 0.15, as depicted by the blue dashed-dotted line shown in Fig.~\ref{fig2}. The lower inhomogeneity is attainable via an intensity down-up operation of the optical lattice, as described in our experimental description.  

The inset of Fig.~\ref{fig2} shows contrast of the Rabi oscillation as a function of pulse area. The measured Rabi oscillation can be fitted by $P^{(0)}_e(A,\gamma) \simeq (1-\cos A \cos (\gamma A))/2$ in the perturbative regime. We can extract the inhomogeneity of the experiment.  
In the inset of Fig.~\ref{fig2}, the maximum excited population of the over-$\pi$ pulse regime is smaller than that of the $\pi$ pulse. The S/N ratio of the excited population is proportional to $\sqrt{(1-p)/p}$ where p is the excitation fraction~\cite{ADLudlowTH}. Therefore it decreases due to the reduced excited population.

The frequency noise of the clock laser and lock-servo error cause the population deviation from the 90\% lock point, therefore induce the unwanted lock point variation which is a one of shift sources. The over-$\pi$ overdriving scheme is more sensitive to the frequency noise of the clock laser with respect to the under-$\pi$ scheme. In the experiment, the additional uncertainty  $\tilde{u}_{lock}$ from the $\pm 6\%$ fluctuation of the 90\% lock point leads to the residual frequency shift change $u_{lock}$ of a $10^{-17}$ level~\cite{A3}. 

We repeat the five independent measurements at the $1.4\pi$ pulse area. The $A=1.4\pi$ pulse induces a collisional frequency shift closest to zero at an inhomogeneity level of 0.15. The shifts and uncertainties of the experiment are given in Table~\ref{table2}. Comparing $\tilde{u}_{col}$ of the $A=1.4\pi$ case with that of the $A=0.6\pi$ case, we show that the over-$\pi$ pulse gives the system a low level of total uncertainty by decreasing the uncertainty from the atomic density fluctuation at the expense of a small $\tilde{u}_{lock}$ increment.
 \begin{table}[ht]
 \caption{ Uncertainty budget of the experiment where $V_{eg}-V_{gg} = -0.47$ Hz, $V_{ee}-V_{gg} = 0.39(V_{eg}-V_{gg})$, $U_{eg}-V_{gg} = -0.35(V_{eg}-V_{gg})$ }
 \centering 
 \begin{tabular}{c c c c c c c c}
 \hline
 \hline
 A & $\gamma$ & $\tilde{\delta}_{col}$ & $\tilde{u}_{col}$ & $\tilde{u}_{lock}$ &   $\tilde{u}_{tot}$& $u_{tot}$ \\
 & &$(\times 10^{-3})$&$(\times 10^{-3})$ & $(\times 10^{-3})$&$(\times 10^{-3})$ & $(\times 10^{-18})$ \\
 \hline
 
$0.6\pi$& 0.15 & 638 & 115 & 12 & 116 & 105\\
$1.4\pi$& 0.15 & 74 & 26 & 19 & 26 & 29\\

 \hline
 \hline
 \end{tabular}
 \label{table2}
 \end{table}
 A small fluctuation below $\pm 1\%$ and a small inhomogeneity below 0.1 provide the total uncertainty $u_{tot}$ less than $5\times 10^{-18}$ when $|V_{eg}-V_{gg}|\leq 0.6$ Hz.

In conclusion, we analyzed the collisional frequency shift in an optical lattice clock interrogated by a Rabi pulse. For the perturbative regime in which most optical lattice clocks are operated, the analytical solution was also obtained. It provides convenient expressions to explore the collisional shift and can be used to extract collision interaction energies in Rabi spectroscopy. Based on our analysis, we propose that an over-$\pi$ pulse combined with a small inhomogeneity enables the cancellation of the total collisional frequency shift. This shows the potential for an optical lattice clock with a $10^{-18}$ uncertainty level with Rabi spectroscopy.

\begin{acknowledgments}
 This work was supported by the Korea Research Institute of Standards and Science under the project entitled Research on Time and Space Measurements, Grant No.14011007.
\end{acknowledgments}


\begin{thebibliography}{collisionalshiftsklee_bib}
\bibitem{A0} For these calculations, it is assumed that the fluctuation of the atomic density $\frac{\delta\rho}{\rho}$ is 18\%, the fluctuation of the pulse area $\delta A$ is on the 0.1\% level, and the fluctuation of the inhomogeneity $\delta\gamma$ is 5\%, all of which are typically obtained in experiments without state-of-the-art components.
\bibitem{A-1} The detuning for a 90\% lock is determined as follows: $\delta \simeq \bar{\Omega}\sqrt{\frac{(1-n)\sin^2(A/2)}{\sin^2(A/2)-A\sin A/4}}$, where $n = 0.9$. When $A=1.4\pi$, $\delta \simeq 0.20 \bar{\Omega}$. Therefore, the perturbation $\bar{\Omega}>>\delta$ can be applied to explain the experimental data.
\bibitem{A3} The residual collisional frequency shift change due to lock point variation of $\pm 6\%$ is written as $\tilde{u}_{lock} \simeq \left|\tilde{\delta}_{col}^{(84\%)}-\tilde{\delta}_{col}^{(96\%)} \right|/2$.
\end{thebibliography}
\end{document}